\documentclass{osa-article}
\journal{oe}
\usepackage{amsmath}
\usepackage{graphicx}
\usepackage{dcolumn}
\usepackage{bm}
\usepackage{epstopdf}
\usepackage{amstext}
\usepackage{setspace}
\usepackage{multirow}

\begin{document}

\title{Formation of narrow atomic lines of Rb in the UV region using a magnetic field}

\author{Ara Tonoyan${}^{a}$, Armen Sargsyan${}^{a}$, Rodolphe Momier${}^{a,b}$, Claude Leroy${}^{b}$, David Sarkisyan${}^{a}$ ${}^{,*}$}

\address{\textit{${}^{a}$ Institute for Physical Research, National Academy of Sciences - Ashtarak 2, 0203, Armenia}}
\address{\textit{${}^{b}$Laboratoire Interdisciplinaire Carnot de Bourgogne, UMR CNRS 6303, Universit\'{e} de Bourgogne, 21000 Dijon, France}
}
\email{\authormark{*}sarkdav@gmail.com}

\begin{abstract}
Magnetically induced (MI) transitions (F${}_{g}$ = 1 $\rightarrow$ F${}_{e}$= 3) of ${}^{87}$Rb D${}_{2}$ line are among the most promising atomic transitions for applications in laser physics. They reach their maximum intensity in the 0.2--2 kG magnetic field range and are more intense than many conventional atomic transitions. An important feature of MI transitions is their large frequency shift with respect to the unperturbed hyperfine transitions which reaches $\sim$12 GHz in magnetic fields of $\sim$ 3 kG, while they are formed on the high-frequency wing of the spectrum and do not overlap with other transitions. Some important peculiarities have been demonstrated for the MI 5S${}_{1/2}$$\rightarrow$ 5P${}_{3/2}$ transitions ($\lambda$=780 nm). Particularly, it was shown that using a nanocell with thickness $L= 100$~nm it is possible to realize 1 $\mu$m-spatial resolution which is important when determining magnetic fields with strong spatial gradient (of $>$~3G /$\mu$m). Earlier, our studies have been performed for 5S${}_{1/2}$ $\rightarrow$ $n$P${}_{3/2}$ transition with $n = 5$, while it is also theoretically shown to be promising for the transitions with $n = 6, 7, 8$ and $9$, corresponding to the transition wavelengths of  420.2~nm, 358.7~nm, 334.9~nm and 322.8~nm, respectively.
\end{abstract}

\section{INTRODUCTION}

A high interest has recently been focused on atomic transitions between ground and excited levels of the hyperfine structures satisfying the conditions $F_e - F_g=\Delta F = \pm 2$ (the probability of these transitions is zero, when the external magnetic field is zero), while the probabilities in a magnetic field increase significantly; for this reason, we refer to these transitions as Magnetically Induced (MI) transitions \cite{1,2}. For the case of $\sigma^{+}$ radiation, the MI transitions $F_g = 1 \rightarrow F_e= 3$ of ${}^{87}$Rb D${}_{2}$ line are among the most promising. The following rule has been found for the intensities of MI transitions: the intensities of magnetically induced transitions with $F_{e}- F_{g} = \Delta F = +2$ are maximal (the number of such MI transitions is also maximal) in case of $\sigma^{+}$ polarized radiation, whereas the intensities of MI transitions with $\Delta$F= -2 are maximal (the number of such transitions is also maximal) in the case of $\sigma^{-}$ polarized radiation. An attractive feature of  $F_{g} = 1 \rightarrow F_{e} = 3$ MI  transitions is their large frequency shift with respect to the unperturbed hyperfine transitions in a magnetic field, which reaches ${\sim}$12 GHz in a magnetic field $B$ ${\sim}$ 3 kG. It is important to note that they are formed on the high-frequency wing of the spectrum and do not overlap with others, which is of practical interest for the study of new frequency ranges.

Previous experiments have been performed for 5S${}_{1/2}$ $\rightarrow$ \textit{n}P${}_{3/2}$ transitions for $n=5$ ($\lambda$=780 nm), while theoretically it is shown in Section 4 to be promising also transitions with $n = 6,7,8$ and 9, the corresponding wavelengths being 420.2 nm,  358.7nm, 334.9 nm and 322.8 nm.

The theoretical model used to calculate the behavior of the Zeeman transitions based on building the magnetic Hamiltonian including the whole hyperfine manifold in the {\textbar}F, m${}_{F}$${>}$ basis is described in detail in [3-6].

\begin{figure}[htb]
\centering
\includegraphics [width=0.8\columnwidth]{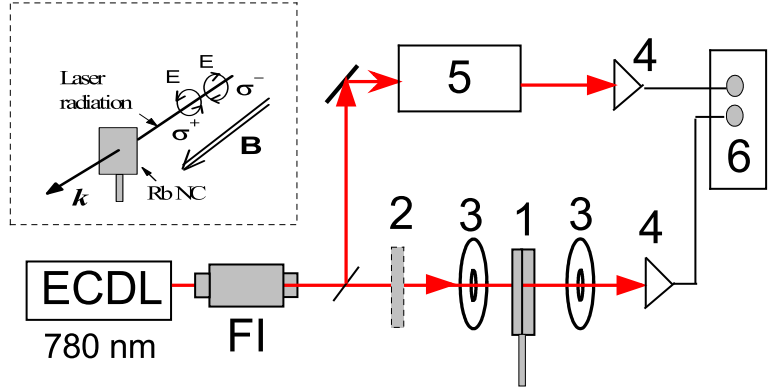}
\caption{Sketch of the experimental setup. ECDL -- diode laser, $\lambda$ = 780nm, FI - Faraday isolator, \textit{1} - Rb nanocell,\textit{2}- quarter-wave plate , \textit{3} are the permanent magnets, \textit{4} - photodetectors, 5-SA unit  for frequency reference,  6\textit{ }-- digital storage oscilloscope, $\textbf{E}$--electric field of laser radiation. Configurations of $\textbf{B}$, $\textbf{E}$ and $\textbf{k}$ are presented in the insets, $\textbf{B} \parallel \textbf{k}$, where $\textbf{k}$ is wave vector of the laser field.}   
\end{figure}

\section{EXPERIMENTAL SETUP}

The nanocell (NC) construction, which consists of 2.4mm-thick windows made of technical sapphire (Al${}_{2}$O${}_{3}$) and a vertical side arm (a sapphire tube filled with Rb) is presented in [7]. The NC is filled with a natural mixture of ${}^{85}$Rb (72.2 \%) and ${}^{87}$Rb (27.8\%). The region of the NC \textit{L} ${\approx}$ $\lambda$/2 is used, where $\lambda$= 780 nm is the transition wavelength of Rb D${}_{2}$ line. The NC operated with a specially designed oven with two ports for laser beam transmission. The temperature of the NC reservoir (which contains the Rb atoms) was ${\sim}$130 ${{}^\circ}$C, but the windows were maintained at a temperature that was 20 ${{}^\circ}$C higher (to avoid condensation of the atomic vapor on the windows ). The experimental setup is depicted on Fig 1. The\textit{ }Rb nanocell (\textit{2}) is located between the strong magnets (\textit{3}). To avoid feedback (reflection from the different optical elements), a Faraday insulator (FI) was used. A part of the laser radiation was directed to the unit (\textit{5}) for frequency reference using the saturated absorption technique. The transmission signal was detected by a photodiode (\textit{4) }and was recorded with a  Tektronix TDS 2014B four-channel storage oscilloscope (\textit{6)}. In order to produce strong magnetic fields two strong permanent magnets were used. Magnetic field \textbf{\textit{B }}is directed along the laser radiation propagation \textbf{\textit{k }}(\textbf{$B$}$\mid$$\mid$\textbf{\textit{k}}). Configuration of \textbf{$B$} and\textbf{\textit{ }}vertical linear polarization (which correspond to circular polarizations $\sigma$${}^{+}$ and $\sigma$${}^{-}$\textbf{\textit{  }}when a longitudinal magnetic field is applied) is shown in the inset of  Fig.1.\textbf{\textit{ }} The$B$-field was measured with a Teslameter HT201 magnetometer.
\begin{figure}[htb]
\centering
\includegraphics [width=0.45\columnwidth]{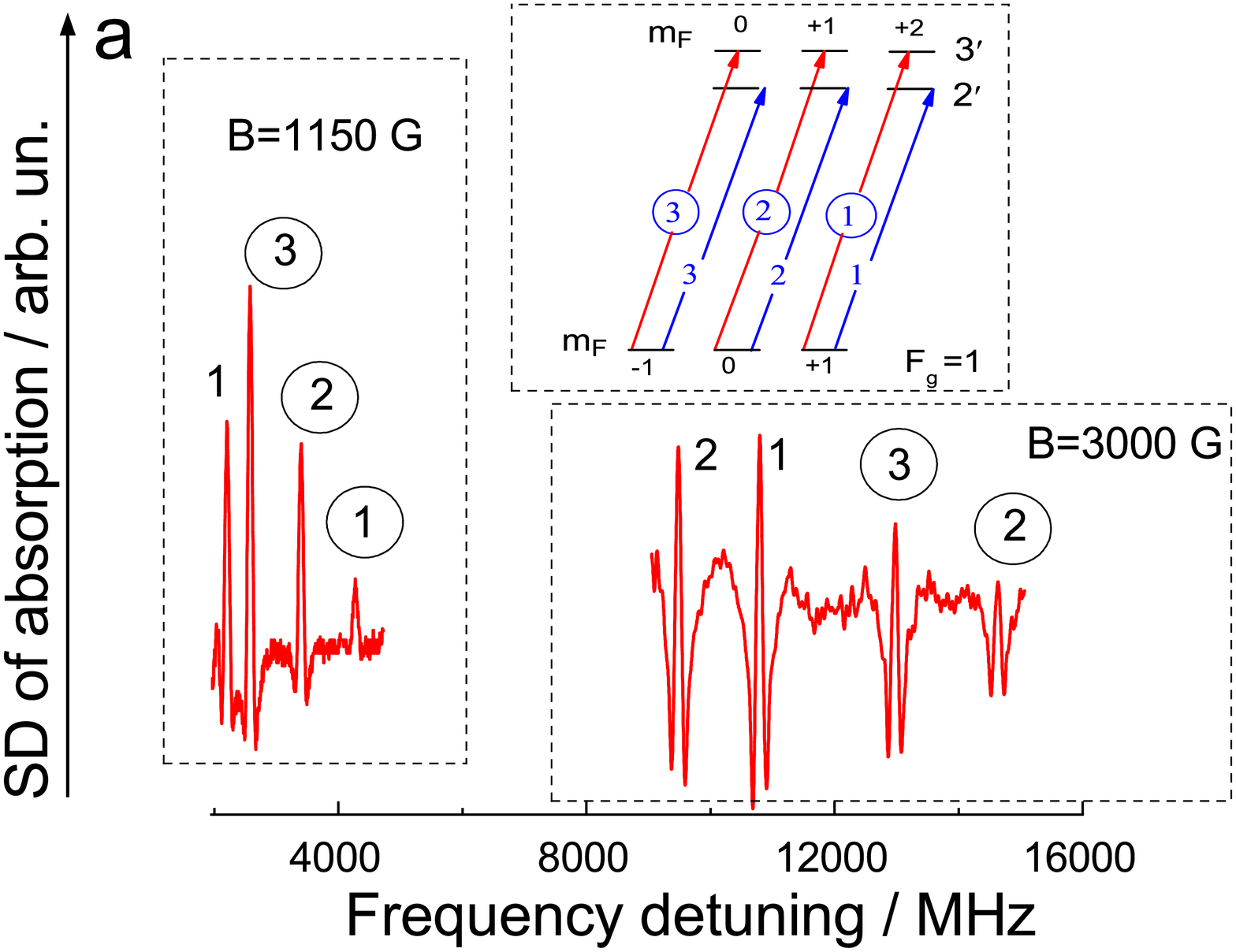}
\includegraphics [width=0.45\columnwidth]{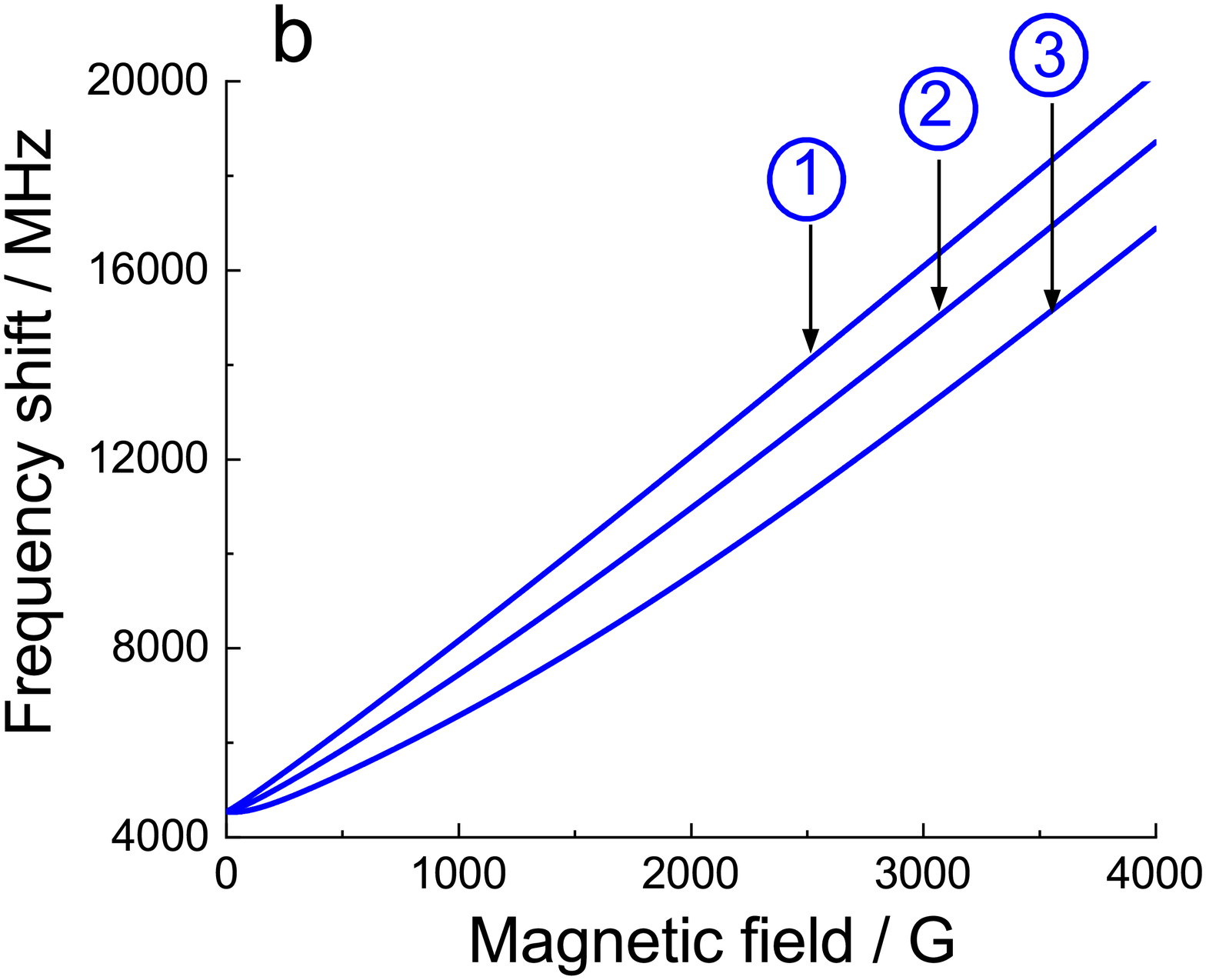}
\caption{(a) The left and right sides show an experimental SD absorption spectrum of atomic transitions \textit{F${}_{g}$ }= 1$\rightarrow$ \textit{F${}_{e}$}=2,3 of the ${}^{87}$Rb D${}_{2}$ line for linearly polarized laser radiation (only spectra for $\sigma$${}^{+}$ are shown) for a $B$ = 1.15 kG and 3 kG, respectively. The inset shows diagram of the \textit{F${}_{g}$ }= 1$\rightarrow$ \textit{F${}_{e}$}=2, 3 transitions. (b) Calculated frequency shift of \textit{F${}_{g}$} = 1$\rightarrow$ \textit{F${}_{e}$}=3 versus $B$-field.}   
\end{figure}

The nanocell (NC) thickness in the direction of  laser beam propagation was chosen to be equal to half the resonant  wavelength of Rb D${}_{2}$ line (\textit{L}=\textit{$\lambda$} /2 = 390 nm). The technique allowing to measure the thickness of an atomic vapor column in the NC is described  in [8]. It has been demonstrated earlier that in this case (so called \textit{$\lambda$}/2 method), narrowing of atomic transitions (lines) in the absorption spectra A($\omega$) of the NC occurs [8,9]. This effect is called Coherent Dicke Narrowing. To obtain further narrowing of the atomic lines, we performed second derivative (SD) of the absorption spectrum A$^\prime$$^\prime$ ($\omega$) [5, 10,11]. In Fig. 2a, the left and right sides show experimental SD absorption spectra of atomic transitions \textit{F${}_{g}$ }= 1$\rightarrow$ \textit{F${}_{e}$}=2, 3 of the D${}_{2}$ line of ${}^{87}$Rb for linearly polarized laser radiation (only spectra for $\sigma$${}^{+}$ are shown) for $B$ = 1.15 kG and 3 kG, respectively. In the case of $B$ = 3 kG the frequency shift of the strongest MI transition \textit{F${}_{g}$}=1, m${}_{F}$=-1$\rightarrow$\textit{F${}_{e}$}=3, m${}_{F}$=0 reaches ${\sim}$9 GHz, with respect of the initial position shown in Fig.2b, while for $B$=4 kG (these MI transitions still have high intensity) it reaches ${\sim}$ 17 GHz. The inset shows diagram of the \textit{F${}_{g}$ }= 1$\rightarrow$ \textit{F${}_{e}$}=2, 3 transitions.

\begin{figure}[htb]
\centering
\includegraphics [width=0.7\columnwidth]{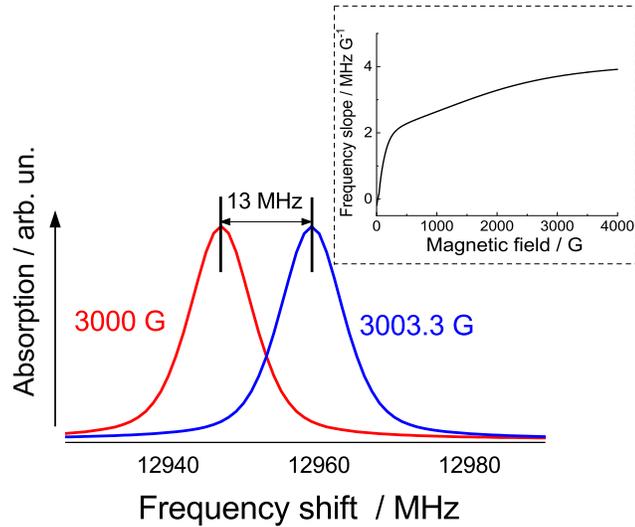}
\caption{Theoretical SD absorption spectra: the red curve has been calculated for $B$${}_{1}$ = 3000 G, and the blue curve has been calculated for $B$${}_{2}$ = 3003.3 G. When moving the NC by 1 $\mu$m (fields $B$${}_{1}$ and $B$${}_{2}$), the frequency shift become 13 MHz which can be measured well. The inset shows the frequency slope of 1$\rightarrow$3$^\prime$ MI transitions.} \end{figure}

\begin{figure}[htb]
\centering
\includegraphics [width=0.7\columnwidth]{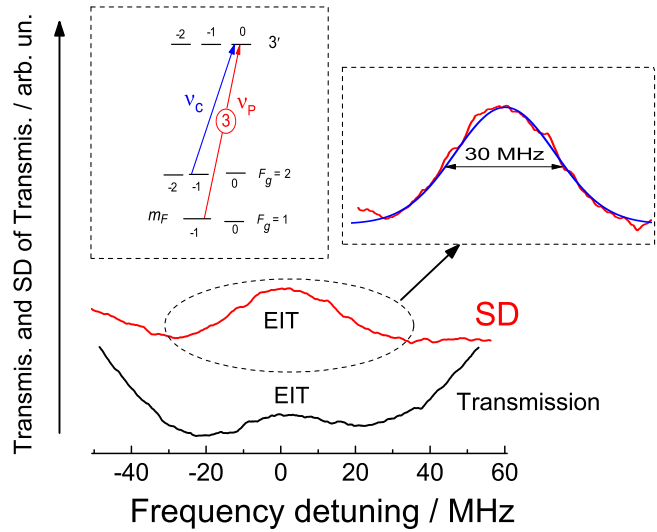}
\caption{The curve labelled ``transmission'' shows an EIT-resonance using a NC with \textit{L}=100 nm, using two diode lasers, the coupling and probe laser power are 30 and 0.1 mW respectively. The temperature of the NC is 180${}^{o}$C. Left inset shows the $\Lambda$-system formed by the probe radiation (the MI transition 1-3$^\prime$) and the coupling radiation 2-3$^\prime$. Right Inset shows the second derivative of the EIT-resonance, which is fitted by the Gaussian curve, the line-width is of around 30 MHz.} \end{figure}
As shown in the works [12,13], an important advantage of the nanocell use is the possibility to measure strong magnetic fields gradients with a high spatial resolution. The paper [14] provides a description of a Stern-Gerlach type deflecting magnet, intended to deflect beams of paramagnetic  nanoclusters, molecules and atoms using a magnetic field with  a gradient of 3.3 G/$\mu$m. Figure 3 shows theoretical SD absorption spectra (red and blue solid lines with FWHM of 10 MHz  and fitted with a Voigt profile) that can be used with the help of the MI transition 3 (see Fig. 2b) of ${}^{87}$Rb to measure such a gradient. If a NC with a thickness \textit{L} = 100 nm is placed on a micrometer stage (magnetic field $B$${}_{1}$ = 3 kG) and shifted by 1$\mu$m with respect to the initial position, $B$ becomes $B$${}_{2}$ = 3003.3G. The red curve has been calculated for B${}_{1}$ = 3000 G, and the blue curve has been calculated for $B$${}_{2}$ = 3003.3 G. As can be seen from Fig.3, when moving the NC by 1 $\mu$m (fields B${}_{1}$ and B${}_{2}$), the transition was shifted of 13 MHz, which can be measured well. The inset of Fig. 3 shows the frequency slope (MHz/G) of 1$\rightarrow$3$^\prime$ MI transitions, which is 3.7 MHz/G for B${\sim}$3 kG and tend to 4 MHz/G at higher $B$-field.

In the work [2] it has been experimentally demonstrated that ${}^{87}$Rb MI transitions 1-3$^\prime$ have been used successfully to form EIT-resonances in $\Lambda$-systems in magnetic field up to 3 kG. For this purpose the NC was used with \textit{L}${\sim}$800 nm. In Fig.4 the curve labeled "transmission" shows a transmission spectrum with an EIT-resonance obtained experimentally using a nanocell with \textit{L}=100 nm in the transmission spectrum, using two independent diode lasers with 1 MHz linewidth. The $\Lambda$-system which is formed by the probe radiation (the MI transition 1-3$^\prime$) and the coupling radiation 2-3$^\prime$ is shown in the left inset. The curve labeled SD is the second derivative of the transmission spectrum, which is well fitted by a Gaussian curve with FWHM 30 MHz, as presented in the right inset. In order to achieve 1 $\mu$m-spatial resolution shown in Fig. 4 it is needed to reduce the EIT-resonance to ${\sim}$10 MHz or less. To implement this, we turn to the works [15-17] in which coherently coupled probe and coupling radiations derived from a single narrow-band laser beam were used. Using a vapor thin cell with \textit{L}=100 $\mu$m they obtained 10-kHz narrow EIT-resonance. This means that using coherently coupled probe and coupling radiations and reducing the thickness \textit{L} by a factor of 1000 (this will increase the collisional linewidth by the same factor), therefore it will be possible to obtain an EIT resonance with a width of at least 10 MHz for \textit{L=}100 nm and to achieve 1 $\mu$m-spatial resolution, as explained in Fig 3a and above.

\section{5S${}_{1/2}$- \textit{n}P${}_{3/2}$ TRANSITION FOR PRINCIPAL QUANTUM NUMBER \textit{n }= 5,6,7,8 AND 9}

\begin{figure}[b]
\centering
\includegraphics [width=0.45\columnwidth]{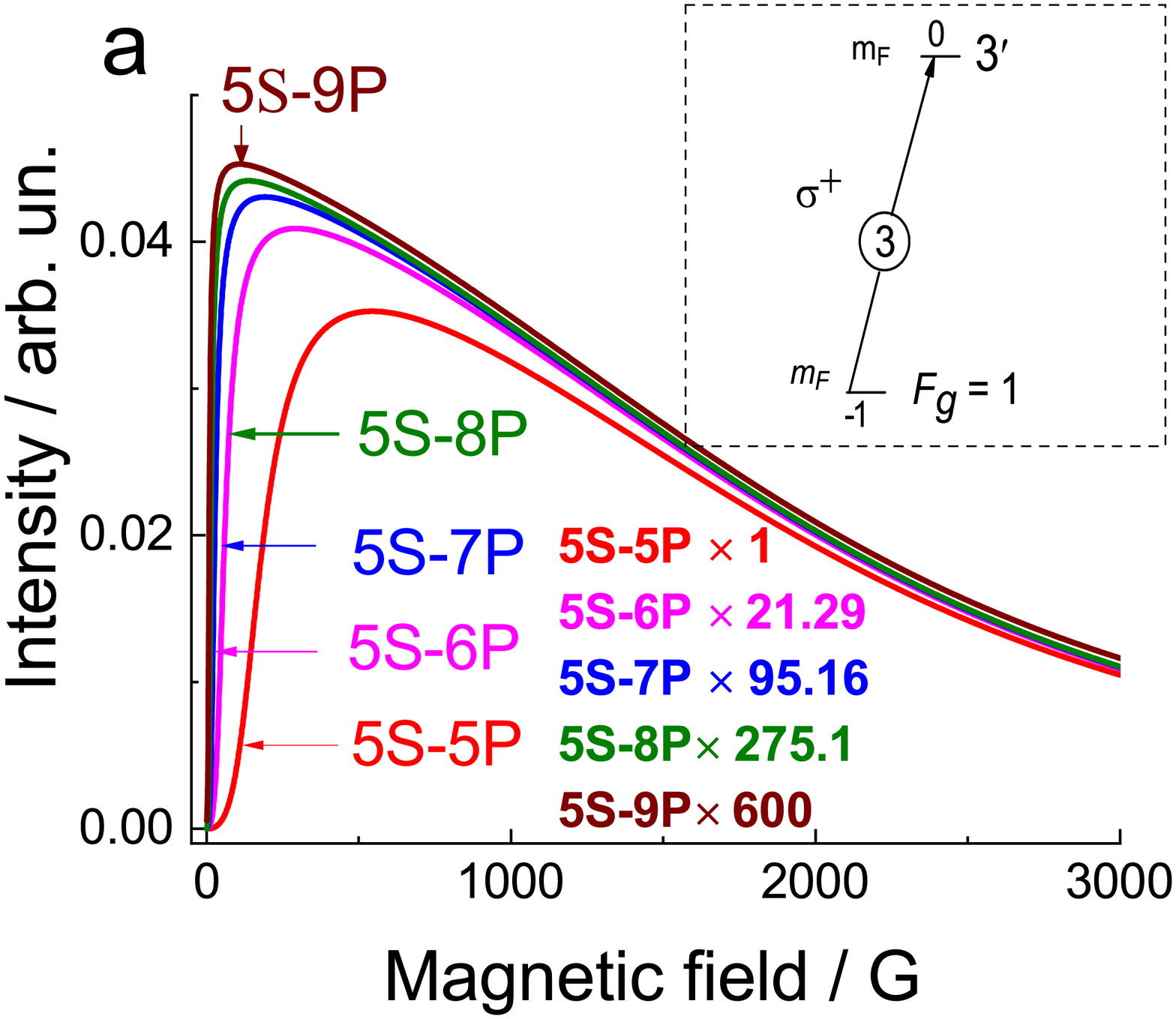}
\includegraphics [width=0.45\columnwidth]{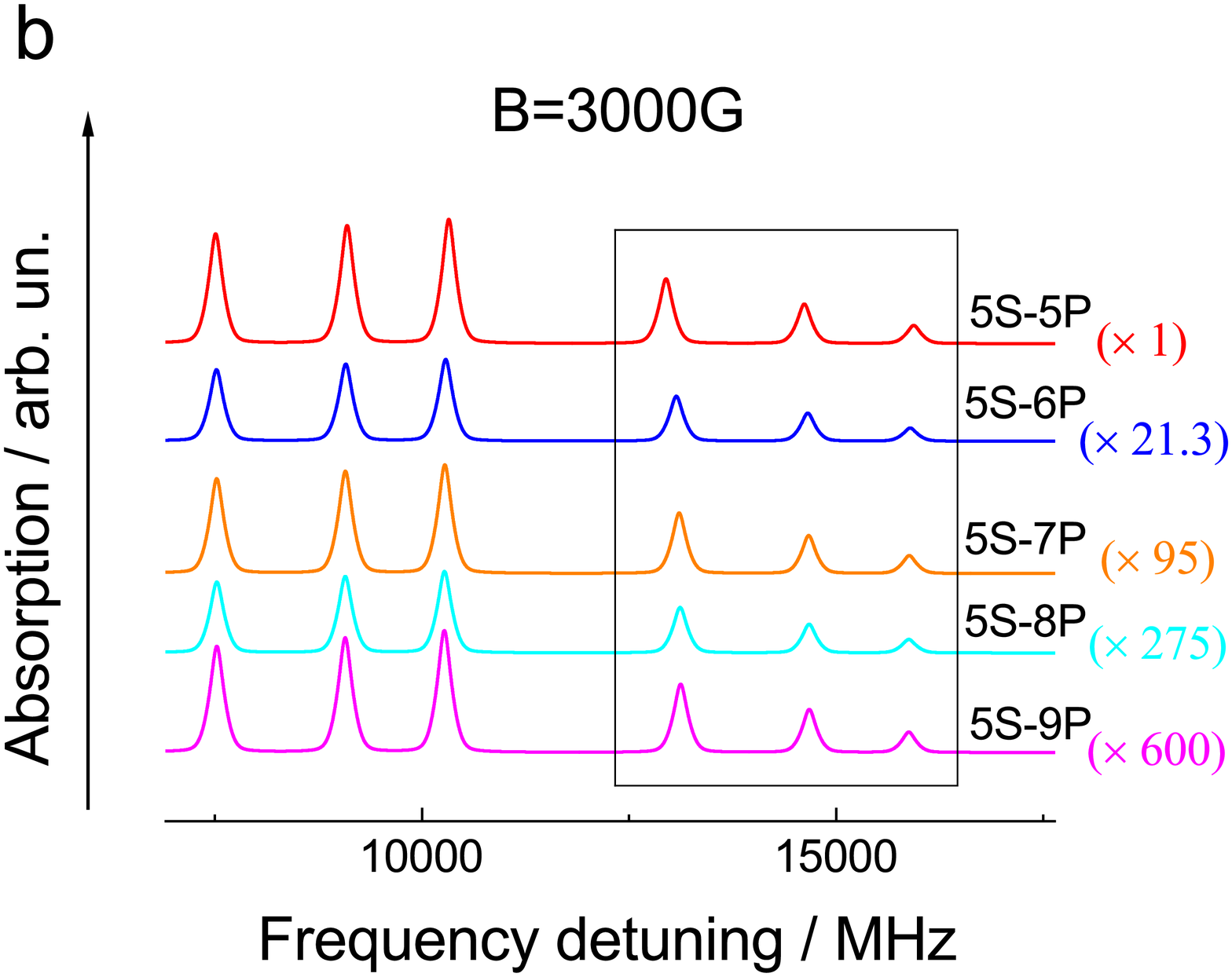}
\caption{(a)  Intensities of  ${}^{87}$Rb , 5S${}_{1/2\ }$- nP${}_{3/2}$, MI \textit{F${}_{g}$}=1, \textit{m${}_{F}$${}_{\ }$}= -1$\rightarrow$\textit{F${}_{e}$}=3, \textit{m${}_{F}$${}_{\ }$}= 0 transition intensity versus $B$-field for the $\sigma$${}^{+\ }$radiation, for principal quantum number \textit{n} = 5,6,7,8,9. Since the intensity of the MI transition decreases with increasing in \textit{n}, the figure indicates by what value the intensity of the corresponding transition is multiplied. (b) Spectra of 5S${}_{1/2\ }$- \textit{n}P${}_{3/2\ }$MI \textit{F${}_{g}$}=1$\rightarrow$\textit{F${}_{e}$}=3, transitions versus $B$-field for the $\sigmaup$${}^{+\ }$radiation, longitudinal magnetic field of $B$ = 3kG is applied. The MI are formed on the high-frequency wing of the spectrum and do not overlap with others.} 
\end{figure}
All previous studies have been done for 5S${}_{1/2}$- \textit{n}P${}_{3/2\ }$transition for \textit{n }= 5 ($\lambda$=780 nm), while theoretically is shown to be promising also transitions with n=6,7,8 and 9, the corresponding laser radiation wavelengths  are 420.2 nm  358.7nm  334.9 nm and 322.8 nm. Numerical simulations of the strongest MI transition for atomic transitions of 5S${}_{1/2}$- \textit{n}P${}_{3/2}$ of ${}^{87}$Rb D${}_{2}$ line, where \textit{n} = 6,7,8,9 while the corresponding laser radiation wavelengths  420.2 nm  358.7nm  334.9 nm and 322.8 nm versus magnetic field are presented in Fig.5a. With an increase in principal quantum number \textit{n}, the magnetic field at which the maximum intensity of the MI transition is reached decreases, which is important for practical applications. Calculations show that for \textit{n} $\geq$ 17 ($\lambda$ $\leq$ 301.5nm) the earth's magnetic field is sufficient for the ``forbidden'' MI transition to become allowed. It's important to note that the usage of a nanocell filled with the Rb vapor with a half-wavelength thickness [2,9] could make it possible to form narrow atomic transitions \textit{F${}_{g}$} = 1 $\rightarrow$ \textit{F${}_{e}$}= 3 in the UV region. Since the probabilities of all 5S${}_{1/2\ }$- \textit{n}P${}_{3/2\ }$transitions decreases with increasing of the principal quantum number \textit{n}, the MI transition remains among the strongest atomic transitions in their group.

\noindent In Fig.5b, spectra of the \textit{F${}_{g}$}=1$\rightarrow$\textit{F${}_{e}$}=3 MI transitions of the 5S${}_{1/2\ }$- \textit{n}P${}_{3/2}$ (\textit{n} = 5,6,7,8,9)${}_{\ }$lines of${}_{\ }$${}^{87}$Rb versus magnetic field for the $\sigma^{+}$${}^{\ }$polarized radiation, for principal quantum number are presented. Longitudinal magnetic field of \textit{B }= 3kG is applied. The MI are formed on the high-frequency wing of the spectrum and do not overlap with others, which is of practical interest in laser spectroscopy.

The frequency shift of MI transitions in strong magnetic fields can reach several tens of gigahertz with respect to the unperturbed hyperfine transitions, which is significantly interesting for potential application of using new frequency ranges, e.g. for laser frequency stabilization at frequencies strongly shifted from the initial transition frequencies of unperturbed atoms [18,19]. Note that in the case of sensitive photo receiver use the formation of the narrow atomic lines could be obtained in a NC at a room temperature [20].

\section{CONCLUSIONS}

\noindent All previous experimental and theoretical studies of the MI transitions of ${}^{87}$Rb, \textit{F${}_{g}$}=1$\rightarrow$\textit{F${}_{e}$}=3 have been done for 5S${}_{1/2\ }$- \textit{n}P${}_{3/2\ }$transition for \textit{n} = 5 ($\lambda$ = 780 nm). In the manuscript an experimental second derivative absorption spectra of atomic transitions \textit{F${}_{g}$} = 1$\rightarrow$ \textit{F${}_{e}$}=2,3 for $\sigma$${}^{+}$ circularly polarized laser radiation for a longitudinal $B$-field of 1.15 kG and 3 kG, respectively are presented which show good spectral resolution, when using a Rb nanocell with \textit{L}${\sim}$ 400 nm  A description is given of a technique based on the EIT process, a nanocell with \textit{L}=100 nm, and a magnetically induced transition, which makes it possible to realize a 1-$\mu$m spatial resolution of a magnetic field with a big gradient.

 Here is theoretically shown the use of the transitions with \textit{n} = 6,7,8 and 9, with the corresponding laser radiation wavelengths  are 420.2 nm, 358.7nm, 334.9 nm and 322.8 nm. The technique based on the nanocell use described above could be also successfully adopted for the UV spectra.

\section*{Funding}
The work was supported by the NATO Science for Peace and Security Project under grant G5794. The work was also supported by the Science Committee of RA, in the frames of project N~1-6/23-I/IPR. 
 
\section*{Disclosures}
The authors declare no conflicts of interest.

\end{document}